# Impact of IT on Higher Education through Continuing Education

**SHAJEE MOHAN B .S.**
**Assistant Professor, C .S .E .D.**
**L.B.S.College of Engineering, Kasaragod, Kerala 671 542.**

**Abstract**

*Information Technology is emerging to be the technology of the 21$^{st}$ century. The paradigm shift from industrial society to information society had already become a reality! It is indeed a high time to think about integrating IT in all facets of education – may it be in secondary or higher education level, or be it in reskilling the employed ones. Reskilling is now a days seriously considered by many corporate firms the ultimate way to educate it's employees as a continuing education scheme to make them stand against the stiff challenges offered by the rapid changes that happening in the IT based high tech industries. Organisations are left with little choice in integrating reskilling in to their development plan. The integration of IT in higher education will itself engender several broader innovations. IT based teaching and learning strategies will open possibilities for designing new curricula and new methods of assessment to meet our educational objectives. This will in addition provide schools with autonomy to use IT resources flexibility to meet the needs of their pupils; school designs will also evolve to maximize the potential for using IT to enhance learning and school administration. IT can be used to promote greater and more efficient communication within the school, amongst schools. It would enhance the effectiveness of educational administration. Ready access to online data and information will also support effective decision-making at all levels. This paper discusses various issues and the need to think about a task force to counter the so-called slowdown and the recession in IT industry. The importance of reskilling as a continuing education programme to make the people aware of the changing trends in IT was also discussed.*

## 1. Introduction

Introduction of Information Technology in various walks of education is in fact the need of the hour. Continuing education or formal education requires careful introduction of IT related techniques to upgrade the standards of teaching and there by improving its effectiveness and efficiency. In order to equip the society with the right knowledge to face the rapid changes that is happening in the technological circles, the IT based teaching is the ultimate solution.

## 2.0. IT Education – The Need of the hour

The integration of IT in education will itself open up several avenues of broader innovations. IT based teaching and learning strategies will open possibilities for designing new curriculum and new methods of assessment to meet our educational objectives.

The Information Technology industry in India grew in the early nineties, when there was a lot of interest among students for a career in the field. This was a time so desperate for the IT workers that it seemed like anyone who could boot a computer could earn a six-figure salary in the US! It was in late 90's that the IT training and education industry received a boost. There was a need to train students in the latest technologies and the formal education sector was not so far completely geared up for the task. Engineering colleges and Universities had old course structure for computer education. Schools in secondary and higher secondary levels not even thought of a master plan to integrate IT in their curriculum. It was at this juncture, the private sector in India jumped in to the arena to fill up the gap. Many firms jumped into reap the revenues out of IT education, but a few of them endured due to the fact that they were not having adequate expertise in this field.

It might seem ironic that most of the developments in the Information Technology first take place in the West before finding their way in to India.

## 3.0. Computer Aided Learning- A Methodology for Continuing Education

New communication and information technologies have become major resources for teaching and learning in higher education. Some of the most cost effective and appropriate ways to use computers and modern technologies is to have close contact between the teachers and the taught.

### 3.1. Web based learning

The wide range of applications that universities and other educational centers have made available on the web is amazing. The latest and the most proficient development in the information revolution is the e – learning through World Wide Web. Now e-mail, computer conferencing and the Internet increase opportunities for students and teachers to converse and exchange work much more speedily than before. Web technology can now enrich learing opportunities like simulation technique with multimedia using latest computer software. It helps the students to develop insight of the subject and to collect information, which may not be immediately available in the library.

Proper instructional strategies must be adopted to make the web based learning more effective. Instructional design is a systematic approach to designing effective instruction. Instructional development provides a process and frame work for systematically planning, developing and adapting instruction based on identifiable learner needs and content requirements.

### 3.2. Web Based Training

Web based training is probably the most powerful advance in training delivery this century. It allows the convergence of different media and also on common platform, the desktop computer. Web based learning is not a total replacement of other training tools or techniques. The delivery of instruction in course using Internet Technology is known as WBT. Web based training uses any combination of texts, graphics, animation, sound, video, or external data banks to present a course of instruction. WBT is advantageous as information and knowledge is available just when it is required. It is more interactive and can send information and receive feedback. Instant updating of information and immediate feedback from users via e-mail is possible. But the drawback is due to the slow bandwidth and slow delivery of interactive material.

### 3.3. Multimedia based learning

The concurrent developments in multimedia, telecommunications and computing technologies, largely driven by consumer electronics markets, have made available convenient tools whose exploitation for education purpose poses creative challenges to academia.

### 3.3.1 Advantages of multimedia instructional tools

The multimedia instructional tools provide an interactive, individualized, self-placed, flexible and motivating learning environment. They are easily adaptable to different learner needs, as for example, to both formal and continuing education. They are beneficial for both students and teachers, the former from the guided learning environment, and the latter by being able to communicate complex ideas more effectively and thus motivating the students.

## 4.0. Continuing IT Education - The Reskilling

Technology is changing at such a rapid ways that even the best of expert's trend to become obsolete with in no time. This trend is even more conspicuous in high-tech industries like IT. Gone are the days when some one could legitimately boast 15 or 20 years of experience in a particular type of furnace, rolling mill or a chemical plant. With average technology life being between 3-5 years, there is a continuous need to upgrade knowledge and skill, of all employees. Similarly, to address the need of more demanding environments and newer technologies. there are many new and better ways of doing the same old business that get introduced almost everyday. Changes in technology and process go together. The old ways of managing projects, people and customers do not working today's markets, and this is yet another reason for reskilling.

Information Technology has been evolving on a daily basis. There is notable change in every aspect of the IT environment. It is indeed evident that the ability of the software industry to sustain its growth will depend upon its ability to integrate the needs of its customers and aspirations of its employees. One this is certain, every one is required to make a change to be effective and make that critical value addition.

The changing environment also increases the pressure to perform on individuals. Considering the present scenario, organizations are left with little choice in integrating reskilling into their development plans. Reskilling should be seen as a "way of life" in organizations, to be effective and add value. It is the only route to keep people and organizations from becoming redundant and obsolete.

### 4.1. The Benefits of Reskilling

" Changing needs and technology," mean just one thing, that most companies will need to train and develop employees on a continual basis. When a company has the opportunities for growth, change or modification, it has broad implications for the training and development function. Such practices will maintain employee's enthusiasm in the job. Training and reskilling can inspire loyalty. A mixed approach to valuing staff by developing skills, providing interesting and motivating work while recognizing their individual contribution, alongside benefits and perks, will mean that you are an employer that employees don't want to leave.

### 4.2. Issues Relating to Reskilling

There are several issues faced during the Reskilling process. Some of them are:
- Reskilling does not occur over night. Signification effort, thought and lead-time are required to do it right. Time must be allocated for this purpose.
- Reskilling will likely need to make compromises in the way things are done and accept new methods and technology
- The set-up costs and demands for continual enhancement means a constant strain on budgets.
- People are reluctant to look at few skill sets. Employee's moral issues arise in all reskilling environments. Managers must invest a significant amount of time to address moral problems

### 5.0. The Recession

The advent of the Internet and e-commerce was an indication of an up search in the IT industry. That gave technology an entirely novel dimension all together. Training institutes, majority of them belongs to the private sector, mean while got busy. The response of students to Java related technologies were over whelming. And in no time, the country was dotted with dotcoms. Incidentally there were also people who entered the field with distorted perceptions. Most of these companies did not have realistic revenue models and failed predictably to meet their revenue targets. The resulted in unanticipated closure of these companies resulting in a downward slide for web professionals. Then came the recession that sent panic waves across the industry. Another fact, which greatly contributed towards this downward trend, was the slow down in the US economy. This has vastly reduced demand for IT professionals all around the world. The September 11 attack on WTC also sent panic waves across the IT industry.

### 5.1. IT task force—The Way

The recession, has however, steadied itself and the exaggerated apprehensions have also been brushed aside. In any case, time had come for people to realize that the IT vision in India was not meant to be confined to just sending professionals to US! Sooner this got cleared the tendency to try out the relatively unexplored areas in the IT industry got stronger. IT education and training institutions have a bigger role to play at this point of time. They need to join hands and take the initiative to overcome the so-called slowdown by forming "IT Taskforce". They should look at the following pointers.
- The IT education centers need to setup a nation-wide and statewide network of educational institutions. These centers should be certified by high quality certification so that a student who studies at any center anywhere in the country is given uniformity in terms of infrastructure, teaching aids and quality education
- Constitute national awards for pioneering work in spreading IT education to the masses
- Initiate IT faculty development programs at the national level and developing new courses and curriculum in IT. A team of consultants, trainers and technologists should be involved in the delivery.
- Integrate IT education from secondary level onwards. The skills required for the future will center on thinking skills, learning skills and communication skills. IT based teaching and learning in the secondary level itself should be one of the strategies for equipping our young with these skills.
- Setting up schools of advanced studies and special research groups in IT.
- Strategic alliances with global majors Microsoft, Oracle, Sun Microsystems etc. can give them continuous feed back on the latest trends prevailing.

- Promote interaction between educational institution and big corporate in the IT industry.
- The government should give financial assistance to IT education centers and the students.
- The Education centers should plan to invest funds for taking the online initiative for developing a separate online training through Internet.
- Single window education model that caters to the career seeking segment for providing guidance, training and placement support has to be powered.

Information Technology is moving fast. It is very important to understand that what do you know doesn't become irrelevant tomorrow. Product renovation and curriculum will need metamorphic changes.

## 6.0. The e-learning initiatives in India

The scope for IT education is limitless, how effectively we use this revolutionary technology is the critical factor. IT can be used as a mass literary and education delivery system for millions of children living in far – flung villages through wired classrooms with a Net teacher. The Internet can provide access to an unlimited storehouse of knowledge on any subject and make remote teaching and distance education a reality. However one should not get carried away by IT hype. IT is just an enabling technology.

One of the first entrants in to the revolutionary IT – driven learning market was the Delhi based NIIT, which introduced it's IT- enabled learning project LEDA (learning through exploration, discovery and adventure) for schools in 1996.since introduction in 1996 over 350 schools and100 other educational firms have experienced LEDA. In the same year they set up an online learning centre-netvarsity.com.

Egurucool entered in the market in 1998. Starting out as a general education portal, the company has moved up the value chain to offer online tutorial courses for ICSE, CBSE and other state board examinations. Egurucool also offers its online services to schools to enable their faculties to plan homework, class work etc.,

Another entrant in to the IT driven and Net learning technology was schoolNet. SchoolNet has trained over 2000 teachers across the country and in the process of expanding their network.

Karnataka was the first state in India to announce its IT policy. Under the Millennium IT policy they have launched several projects to utilize the power of IT to accelerate the overall development of the state. Their most ambitious InfoTech initiative is IT in 1000 schools. Five-year computer literacy project estimated to cost the government Rs.190 crore, will focus on 1000 government run secondary schools spread across the state. Another initiative in this direction is yuva.com, a scheme under which 225 computer training centers will be established for rural youth and an IT literacy project among all of the state's 77 engineering colleges, 100 polytechnics, 150 ITIs and 300 other colleges

Tamilnadu government introduced a computer literacy programme in 666 government schools in the state. The Chennai Municipal Corporation has launched its Virtual classroom programme- an initiative under which all corporation schools in the city are linked via the Internet and offer web-based learning.

The Information Kerala mission is the prestigious project conceived by the government of Kerala to impart the benefits of Information technology to the rural masses. Under this programme the District, Block and Grama Panhayaths were provided with computer systems and Internet facilities to make them aware of the latest trends in this field. This is a humble initiative to incorporate e- governance in future. Under this scheme the government have laid plans to tie up with global giants like Microsoft to provide the software and multimedia support to the states primary and secondary schools to make the e-learning a reality. Through this ambitious project Kerala, the only state in India who achieved 100% literacy, aims to attain 100% computer literacy.

## 7.0. Conclusion

There is no end to technological advancement. Up trends and down trends is quite natural and Information Technology industry is set for a great innings ahead. We should aim for a future where every child will be able to enhance his learning through an IT enriched curriculum and school environment. By integrating IT in education we should be able to achieve our goals like enhancing linkage between schools and world around it, so as to expand and enrich the learning environment, encourage creative thinking, life long learning and social responsibility generate innovative avenues in education and promote administrative and management excellence in the education system to meet the challenges of 21$^{st}$ century.

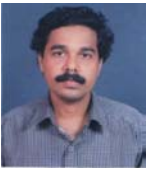
**Mr. SHAJEEMOHAN B. S.** is an Assistant Professor (on deputation) in Computer Science and Engineering Department, L. B. S. College of Engineering, Kasaragod and presently is the Head of Department of CSE. He took his B-Tech from College of Engineering, Thiruvanathapuram in the year 1989 and took his MTech in Computer Science and Engineering from V.T.U. in the year 2000. He has been working in various Premier Engineering Institutions in Kerala since 1991. He is a life member of Indian Society for Technical Education. His research interests are in Parallel processing, Data compression algorithms.